\documentclass[11pt,twoside]{article}

\usepackage{baaa}

\usepackage{graphicx}
\usepackage{subfigure}
\usepackage{amssymb}
\usepackage[spanish,activeacute,english]{babel}
\usepackage[latin1]{inputenc}
\usepackage{verbatim}
\usepackage{amsmath}
\usepackage{amsfonts}
\usepackage{amssymb}
\usepackage[colorlinks=true,dvips]{hyperref}
\usepackage{natbib}

\begin{document}
%%%%%%%%%%%%%%%%%%%%%%%%%%%%%%%%%%%%%%%%%%%%%%%%%%%%%%%%%%%%%%%%%%%%%%%%%%
%%%% SELECCIONE EL IDIOMA EN QUE SE ESCRIBE EL ARTÍCULO:              %%%%
%\myselectspanish
\myselectenglish
%%%%%%%%%%%%%%%%%%%%%%%%%%%%%%%%%%%%%%%%%%%%%%%%%%%%%%%%%%%%%%%%%%%%%%%%%%
\vskip 1.0cm
\markboth{De Rossi et al.}%
{}

\pagestyle{myheadings}
%%%% DESCOMENTE LA LINEA QUE DESCRIBE EL CARACTER DE SU TRABAJO       %%%%
\vspace*{0.5cm}

%\noindent TRABAJO INVITADO 
\noindent PRESENTACIÓN ORAL
%\noindent PRESENTACIÓN MURAL
%\noindent RESUMEN 

\vskip 0.3cm
\title{Evolution of the gas kinematics of galaxies in cosmological simulations}

%\title{ Template paper for publication in the Bulletin of the 
%Argentinian Astronomical Association with instructions for the use of 
%\LaTeX{}}

\author{M.E. De Rossi$^{1,2,3}$ \& S.E. Pedrosa$^{1,2}$}

\affil{%
  (1) Instituto de Astronomía y Física del Espacio (CONICET-UBA)\\ 
  (2) Consejo Nacional de Investigaciones Cient\'ificas y T\'ecnicas, CONICET, Argentina (derossi@iafe.uba.ar)\\
  (3) Facultad de Ciencias Exactas y Naturales, Universidad de Buenos Aires, Ciudad Aut\'onoma de Buenos Aires, Argentina\\
}

\begin{abstract} 
We studied the evolution of the gas kinematics of galaxies by performing hydrodynamical simulations 
in a cosmological scenario.  We paid special attention to the origin of the scatter of the 
Tully-Fisher relation and the features which could be associated with mergers and interactions.
We extended the study by De Rossi et al. (2010) and analysed their whole simulated sample which 
includes both, gas disc-dominated and spheroid-dominated systems.
We found that mergers and interactions can affect the rotation curves directly or indirectly inducing
a scatter in the Tully-Fisher Relation larger than the simulated evolution since $z \sim 3$.
In agreement with previous works, kinematical indicators which combine the rotation velocity 
and dispersion velocity in their definitions lead to a tighter relation. 
In addition, when we estimated the rotation velocity at
the maximum of the rotation curve, we obtained the best proxy for the potential well regardless
of morphology.
\end{abstract}

\begin{resumen}
Estudiamos la evolución de la cinemática del gas en galaxias realizando simulaciones hidrodinámicas
en un escenario cosmológico.  Prestamos especial atención al origen del {\it scatter} de la relación
de Tully-Fisher y los rasgos que podrían ser asociados a fusiones e interacciones.
Extendimos el estudio de De Rossi et al. (2010) y analizamos su muestra simulada completa, la cual
incluye tanto sistemas dominados por discos como esferoides de gas.  Encontramos que las fusiones e
interacciones pueden afectar las curvas de rotación directa o indirectamente induciendo
un {\it scatter} en la relación de Tully-Fisher mayor que la evolución simulada desde $z \sim 3$.
En acuerdo con trabajos previos, los indicadores cinemáticos que combinan la velocidad de rotación y la
velocidad de dispersión en sus definiciones conducen a una relación más estrecha.
Más aún, cuando estimamos la velocidad de rotación en el máximo de la curva de rotación, obtenemos
el mejor sustituto para el pozo de potencial independientemente de la morfología.
\end{resumen}

\section{Introduction}
\label{intro}

Recent observational and theoretical works have suggested that the Tully-Fisher relation (TFR) can
be generalized to include dispersion dominated galaxies (e.g. Weiner et al. 2006; Kassin et al. 2007;
Covington et al. 2010; Vergani et al. 2012).  In this context, a new kinematical indicator ($s_K$) which
combines the rotation ($V_{\rm rot}$) and dispersion velocity (${\sigma}$) is used:

\begin{equation}
s_K^2 = K \ . \ V_{\rm rot}^2 + {\sigma}^2 ,
\end{equation}

where $K$ is a constant $\le 1$.  In particular, Kassin et al. (2007) found that the use of $s_{0.5}$ leads
to a unified correlation between mass and velocity for galaxies of all morphological types
in their sample, which included merging and distubed systems.
By performing pre-prepared merger simulations, Covington et al. (2010) also reported a reduction
in the scatter of the TFR when using $s_{0.5}$. These authors claimed that the scatter in the TFR is closely
related with mergers and interactions.

In this work, we used cosmological simulations to study the kinematics of surviving 
gaseous discs in galaxies of all morphological types.  We derived the simulated TFR
and determined which kinematical indicator is the best proxy for the potential well in
these simulations.

\section {Numerical simulations and galaxy sample}
We performed numerical simulations consistent with the concordance $\Lambda$-CDM universe
with $\Omega_m =0.3, \Omega_\Lambda =0.7, \Omega_{b} =0.045$, a normalisation of the power
spectrum of ${\sigma}_{8} = 0.9$ and $H_{0} =100 \, h$ km s$^{-1} \ {\rm Mpc}^{-1}$ with  $h=0.7$. 
These simulations were performed by using the chemical code GADGET-3 (Scannapieco et al. 2008), 
which includes treatments for metal-dependent radiative cooling, stochastic star formation, 
chemical enrichment and  supernovae feedback.   
The simulated volume corresponds to a cubic box of a comoving 10 Mpc $h^{-1}$ side length. 
The masses of dark matter and initial gas-phase particles are 
$5.93 \times 10^6 M_{\odot} h^{-1}$ and  $9.12 \times 10^5 M_{\odot} h^{-1}$, respectively.

For each simulated galaxy, we determined the gas-phase rotation curve and measured
the rotation, dispersion and circular velocity at different radii.  In particular,
we determined the radius ($R_{\rm max}$) where the rotation velocity is maximum
and defined this velocity as $V_{\rm max}$.
Unless otherwise specified, all simulated properties were calculated inside the
baryonic radius ($R_{\rm bar}$), defined as the one which encloses 83\% of the baryonic
mass of a galactic system.
A detailed description of this analysis and more information about 
these simulations can be found in De Rossi et al. (2010, 2012).

\section{Results and discussion}

As shown in De Rossi et al. (2010, 2012),  these simulations predict a tight
TFR for disc galaxies that is in very good agreement with recent observations (e.g. Reyes et al. 2011).
However, when including dispersion dominated systems, the scatter in the TFR tends to increase
with maximum velocity variations in the range [0.4, 0.5] dex.  Hence, the scatter becomes larger
than the mean velocity evolution since $z \sim 3$ ($\sim 0.1$ dex).
Consistently with previous works, the analysis of the merger trees of simulated
systems indicates that this scatter is mainly caused
by mergers and interactions which drive turbulent and disordered motions of gas 
and also lead to other processes as star formation, gas infall and outflows.
These mergers and interactions alter the morphologies of the remnant galaxies
and strongly infuence the evolutionary tracks of the systems along the TFR-plane.
In particular,  De Rossi et al. (2012) found that the tracks given by $V_{\rm rot}$
and $V_{\rm circ}$ can significantly depart during merger events.
In this context, it is worth analysing if there is a kinematical indicator which
can be considered a good proxy for the potential well regadless of the galaxy morphology
and, in that case, if this kinematical indicator can generate a tighter and unified TFR 
for galaxies of all morphology types,
as it was reported in previous works.   

\begin{figure}[!ht]
  \centering
  \includegraphics[width=0.51\textwidth]{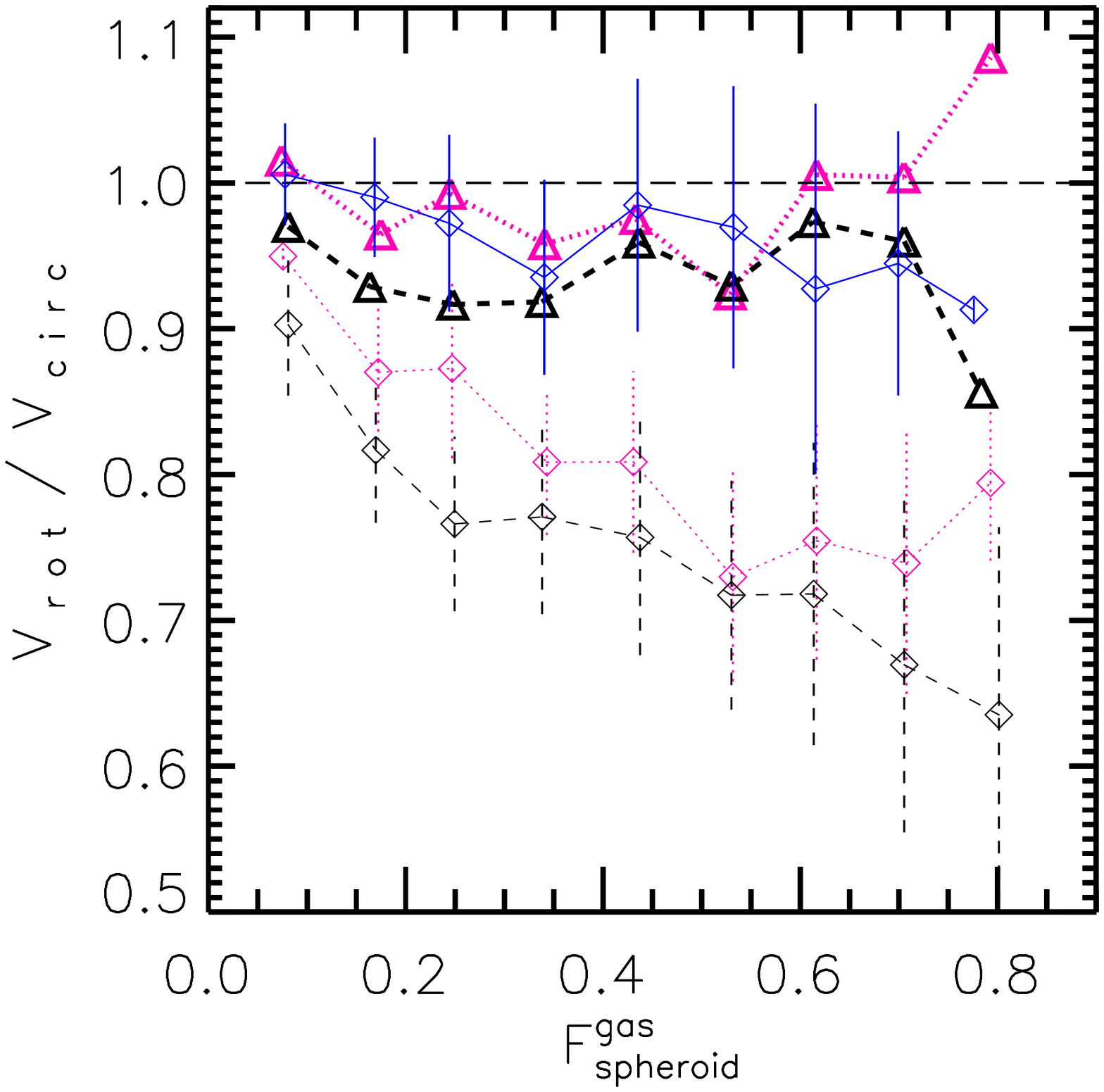} \hspace{-0.55cm}
  \includegraphics[width=0.51\textwidth]{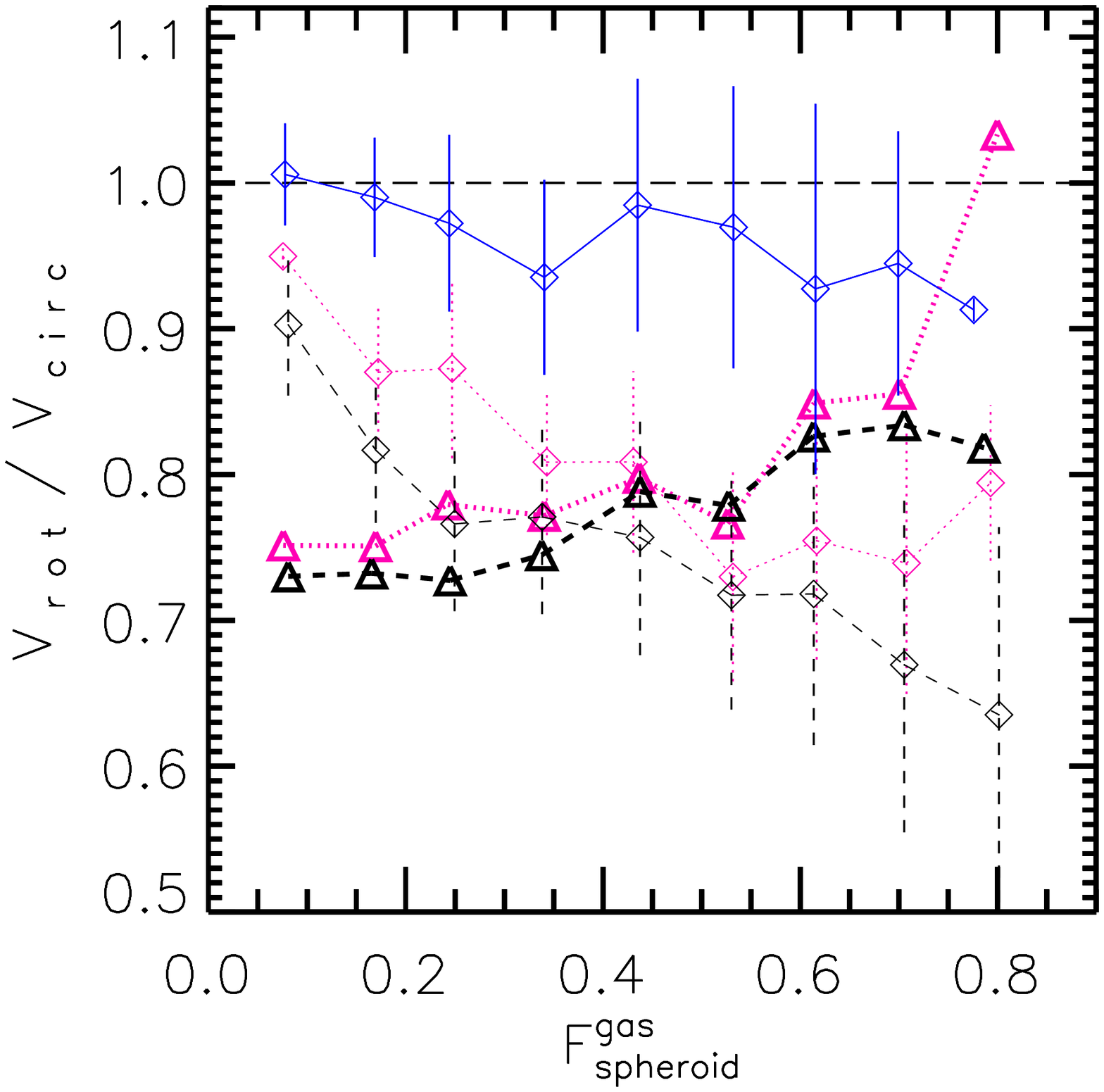}
  \caption{
$V_{\rm rot} / V_{\rm circ}$ (diamonds) 
as a function of the gas-fraction in the spheroidal component of simulated galaxies and at different
radii: $R_{\rm max}$ (solid blue lines), $R_{\rm bar}$ (dotted pink lines) and
$1.5 R_{\rm bar}$ (black dashed lines).  
The curves represent the mean relations while the vertical lines 
correspond to the standard deviations.  
We also show with triangles the mean relations obtained when estimating $s_{1.0} / V_{\rm circ}$ 
(left panel) and $s_{0.5} / V_{\rm circ}$ (right panel)
at $R_{\rm bar}$ (dotted pink lines) and $1.5 R_{\rm bar}$ (black dashed lines).
}
  \label{fig:fig1}
\end{figure}

In Fig. \ref{fig:fig1}, we can appreciate $V_{\rm rot} / V_{\rm circ}$ (diamonds) 
as a function of the gas-fraction in the spheroidal component ($F^{\rm gas}_{\rm spheroid}$) of simulated galaxies 
and at different radii: $R_{\rm max}$ (solid blue lines), $R_{\rm bar}$ (dotted pink lines) and
$1.5 R_{\rm bar}$ (black dashed lines).
The curves represent the mean relations while the vertical lines 
correspond to the standard deviations.
We also show with triangles the mean relations obtained when estimating $s_{1.0} / V_{\rm circ}$
(left panel) and $s_{0.5} / V_{\rm circ}$ (right panel) at $R_{\rm bar}$ (dotted pink lines) and $1.5 R_{\rm bar}$ (black dashed lines).

It is clear from the left panel of Fig. \ref{fig:fig1} that, when the disc dominates the gas-phase (i.e. $F^{\rm gas}_{\rm spheroid} < 0.2$),
$V_{\rm rot}$ and $s_{1.0}$ are good proxies for the circular velocity.  
On the other hand, for gas-spheroid-dominated systems, $V_{\rm rot}$ is a good proxy for the potential well only 
if estimated at the maximum of the rotation curve.
At larger radii ([$R_{\rm bar}, 1.5 R_{\rm bar}$]), $V_{\rm rot}$ tends to understimate $V_{\rm circ}$, 
on average.  However, when combining $V_{\rm rot}$ and ${\sigma}$ in the definition of the kinematical
indicator $s_{1.0}$, a good tracer for the potential well is obtained at large radii, 
in agreement with previous works.  
It is worth noting that in these simulations the best proxy for $V_{\rm circ}$ at large radii
is $s_{1.0}$, while $s_{0.5}$ is used more frequently in the literature.  We verified that $s_{0.5}$
can reduce the scatter of the simulated TFR but as shown in the right panel of Fig. \ref{fig:fig1}, $s_{0.5}$ tends to underestimate $V_{\rm circ}$.
This issue is discussed in detail in De Rossi et al. (2012).

\section{Conclusions}
We analysed the gas kinematics of galaxies in a $\Lambda$-CDM cosmology by using
hydrodynamical simulations.  We obtained a TFR in good agreement with observational works
when using only disc-like systems.  However, when including dispersion-dominated galaxies,
the scatter of the TFR significantly increases.
We found that this scatter is strongly related with galaxy interactions and merger events
that can significantly disturbed the rotation curve of simulated systems, specially at large
radii.  Our simulations predict that this scatter can be reduced by combining 
$V_{\rm rot}$ and ${\sigma}$ in the definition of the kinematical indicator.  In particular,
we obtained that $s_{1.0}$ does not only reduce the scatter but also seems to be a good tracer of the potential 
well for all morphological types, at least in these simulations.
In the inner part of simulated galaxies, 
we found that the best proxy for $V_{\rm circ}$ is
the maximum rotation velocity $V_{\rm max}$, regardless of morphology.

More details about this work can be found in De Rossi et al. (2012).

\begin{acknowledgements}
We thank the anonymous referee for his/her useful comments that helped 
to improve this article.
We acknowledge support from the  PICT 32342 (2005),
PICT 245-Max Planck (2006) of ANCyT (Argentina), PIP 2009-112-200901-00305 of
CONICET (Argentina) and the L'oreal-Unesco-Conicet 2010 Prize.
Simulations were run in Fenix and HOPE clusters at IAFE.
\end{acknowledgements}

\begin{referencias}

\reference Covington, M.~D., et al.\ 2010, \apj, 710, 279

\reference de Rossi, M.~E., Tissera, P.~B., \& Pedrosa, S.~E.\ 2010, \aap, 519, A89

\reference De Rossi, M.~E., Tissera, P.~B., \& Pedrosa, S.~E.\ 2012, \aap, 546, A52 

\reference Kassin, S.~A., et al.\ 2007, \apjl, 660, L35

\reference Reyes, R., Mandelbaum, R., Gunn, J.~E., Pizagno, J., \& Lackner, C.~N.\ 2011, \mnras, 417, 2347 

\reference Scannapieco, C., Tissera, P.~B., White, S.~D.~M., \& Springel, V.\ 2008, \mnras, 389, 1137 

\reference Tully, R.~B., \& Fisher, J.~R.\ 1977, \aap, 54, 661

\reference Vergani, D., Epinat, B., Contini, T., et al.\ 2012, \aap, 546, A118 

\reference Weiner, B.~J., et al.\ 2006, \apj, 653, 1027

\end{referencias}

\end{document}